\documentclass[nofootinbib, 10pt,superscriptaddress]{revtex4-2}
\usepackage{graphicx}
\usepackage[T1,T2A]{fontenc}
\usepackage[utf8]{inputenc}
\usepackage[english]{babel}
\usepackage{amsmath}
\usepackage{amssymb}
\usepackage{perpage}
\MakePerPage{footnote}
\usepackage{caption}
\usepackage{comment}
\usepackage[lofdepth,lotdepth]{subfig}
\usepackage{float}
\usepackage{multirow}
\usepackage{stackengine}[2013-10-15]
\usepackage{dutchcal}
\usepackage[normalem]{ulem}
\usepackage{xcolor}

\addto\captionsenglish{}

\begin{document}

\preprint{INR-TH-2025-008}



\title{ Generalized Models for Spinning Field Lumps on Plane }

\author{Yulia Galushkina}
\affiliation{Institute for Nuclear Research of RAS, prospekt 60-letiya Oktyabrya 7a, Moscow, 117312}

\author{Eduard Kim}
  \affiliation{Institute for Nuclear Research of RAS, prospekt 60-letiya Oktyabrya 7a, Moscow, 117312}
  \affiliation{Moscow Institute of Physics and Technology, Institutsky lane 9, Dolgoprudny, Moscow region, 141700}

\author{Emin Nugaev}
\affiliation{Institute for Nuclear Research of RAS, prospekt 60-letiya Oktyabrya 7a, Moscow, 117312}

\author{Yakov Shnir}
\affiliation{BLTP JINR, Joliot-Curie St 6, Dubna, Moscow region, 141980}

\begin{abstract}
   We study planar non-topological solitons in models with nonlinear potentials that are bounded from below. 
   These models provide consistent completion for the classical consideration at any energy scale. 
   The properties of our solutions indicate the kinematical stability, which is unachievable in the previously studied model with negative quartic self-interaction. 
   Remarkably, our generalization preserves restoration of the full Schr\"{o}dinger symmetry at low energies, including scale invariance (dilatation) and special conformal symmetry. Our numerical calculations and analytical approximations demonstrate that the details of non-relativistic regime are defined by the lowest nonlinear $U(1)$-invariant term.
\end{abstract}

\maketitle

\section{Introduction}

Various nonlinear physical systems support non-topological solitons, which represent localized field configurations with finite energy. One of the simplest examples is given by Q-balls, stable isospinning non-topological solitons that can exist in a general class of scalar field theories with a global $U(1)$ symmetry and an appropriate self-interaction potential \cite{Rosen:1968mfz,Friedberg:1976me,Coleman:1985ki}, for a review see, e.g. \cite{Lee:1991ax,Radu:2008pp,shnir2018topological,Nugaev:2019vru}. 
The Q-balls carry a Noether charge $Q$ associated with this symmetry, it is proportional to the angular frequency of the complex boson field and represents the boson particle number of the configurations. In other words, Q-balls can be considered as condensates of a large number of the field quanta, which correspond to extrema of the effective energy functional for a fixed value of the charge $Q$ \cite{Enqvist:2003zb}. Q-balls have attracted considerable attention in recent years; in particular, it was suggested that they may be formed in a primordial phase transition contributing to various cosmological scenarios \cite{Frieman:1989bx,Kusenko:1997hj} and baryogenesis \cite{Affleck:1984fy,Enqvist:1997si}. Q-balls can be considered as promising  candidates for dark matter (DM) \cite{Kusenko:1997si,Mielke:2002bp,Chen:2020cef}.

Q-ball may exist in various dimensions, in particular the $(2+1)$-dimensional Q-balls \cite{Battye:2000qj,Axenides:1999hs,Volkov:2002aj} are important in the context of condensed-matter \cite{Bunkov:2007fe}. In the non-relativistic limit, which is relevant for effective field theories of ultracold atoms, nonlinear optics, and certain planar condensed-matter systems, the dynamics of complex scalar fields is governed by nonlinear Schr\"{o}dinger equation (NLSE), or Gross-Pitaevskii equation \cite{Enqvist:2003zb}. Such theories naturally support non-topological solitons stabilized by particle-number conservation, furthermore, there are spinning or rotating Q-balls with nonzero angular momentum through a phase winding around the core of the soliton. It was pointed out that a symmetry group of a non-relativistic nonlinear planar theory can be more complex than the usual Galilei group, it can be enhanced to the Schr\"{o}dinger group, which includes dilatations and special conformal transformations \cite{deKok:2008ge}. Recently, we considered this phenomenon in the $(2+1)$-dimensional Gross-Pitaevskii model with the quintic term \cite{Galushkina:2025hkw} and in a theory of a complex scalar field with quartic self-interaction \cite{Galushkina:2025yce}. In particular, contemporary research on dark matter \cite{Brax:2025uaw} considered as vortices in $(2+1)$ dimensions showed the importance of details of the non-relativistic regime for DM dynamics.

In this work, we investigate the existence, structure, and properties of spinning non-topological solitons in $(2+1)$ dimensions within a generic class of non-relativistic scalar field theories. A well-known Lorentz-invariant $(-\lambda|\phi|^{4})$ model has a potential that is unbounded from below and possesses non-topological solitons that are both kinematically and linearly unstable. In this scope, ultraviolet (UV) completion of this theory and the properties of non-topological solitons within UV-completed models are of interest. For this purpose, in this work we consider potentials bounded from below, which effectively contain $(-\lambda|\phi|^{4})$. We also study the generalization of quartic potential within the framework of renormalizable Friedberg-Lee-Sirlin theory \cite{Friedberg:1976me}. In Sec.\ref{section 1}, we overview the analysis of the non-relativistic limit in the model with quartic self-interaction. In addition, a discussion of the properties of relativistic solitons is also given in this section. Then, we introduce our first UV-completed model in Sec.\ref{section 2}. This model is constructed by adding a higher-order $U(1)$ term to the Lagrangian of the original theory. In this model, we have established the region of parameters that allow for stable non-topological solitons. Another attempt to construct the UV completion of the original theory is given in Sec.\ref{section 3}. In this section, we consider the Friedberg-Lee-Sirlin (FLS) model, where a real scalar field is added. Initially, we show that the theory $(-\lambda|\phi|^4)$ is a low-energy EFT for the FLS model using the effective potential method. Next, we study and compare the integral characteristics of non-topological solitons in both the FLS model and our EFT.

\section{Non-topological solitons in $(2+1)$-dimensional conformal model}\label{section 1}

In order to illustrate our approach to non-topological solitons in $(2+1)$ dimensions \cite{Galushkina:2025yce}, let us consider the non-relativistic scalar theory with quartic self-interaction. The corresponding Lagrangian\footnote{We use natural system of units $\hbar=c=1$. The parameters of the theory are of units of mass $M$.} is
\begin{equation}\label{NR lagrangian}
    \mathcal{L}_{NR} = i\psi^{\ast}\dot{\psi} - \frac{1}{2m}\nabla\psi^{\ast}\nabla\psi + \frac{\lambda}{8m^{2}}\left(\psi^{\ast}\psi \right)^{2},
\end{equation}
where \([m] = M\), \([\lambda] = M\), \(\lambda > 0\). This Lagrangian results in the nonlinear Schr\"{o}dinger equation.

The theory (\ref{NR lagrangian}) is invariant under space and time transformations of Schr\"{o}dinger group and global $U(1)$ symmetry. Among the symmetries of the Schr\"{o}dinger group, we highlight scale invariance (dilatations) and special conformal symmetry. Note that for quartic self-interaction these symmetries are exact and unbroken only in $(2+1)$ dimensions. 

Due to the U(1)-invariance and an attractive nature of quartic nonlinear term in Lagrangian (\ref{NR lagrangian}), this model supports non-topological solitons, or Q-tubes. Indeed, let us consider the following ansatz $\psi(t,\vec{x}) = e^{i\mu t}e^{in\theta}h(r)$, where \(\mu \ll m\) is a frequency parameter, \(n\) is a winding number and \(r = \sqrt{x^i x^i}\). Thus, we obtain the equation of motion
\begin{equation}
    h^{''}(r)+ \frac{h^{'}(r)}{r}-\frac{n^2}{r^2}h(r)=2m\mu h(r)-\frac{\lambda}{2m}h^{3}(r).
\end{equation}
One can notice that this equation allows for the scaling
\begin{equation}
\label{NR_sc}
 \bar{r} =r\sqrt{2m\mu}~ , \quad \bar{h} = h\sqrt{\frac{\lambda}{2m\mu}},
\end{equation}
thus, it is rewritten in a way that it becomes independent of both \(\mu \text{ and } \lambda\)
\begin{equation}
    \bar{h}^{''}(\bar{r})+ \frac{\bar{h}^{'}(\bar{r})}{\bar{r}}-\frac{n^2}{\bar{r}^2}\bar{h}(r)= \bar{h}(\bar{r})-\frac{1}{2m}\bar{h}^{3}(\bar{r}).
\end{equation}
This equation may be solved numerically, so one can find the conformal Q-tube solutions. 

These solutions posses peculiar features due to their scale invariance and conformal symmetry. To see that, let us consider their integral characteristics. The Hamiltonian of the model (\ref{NR lagrangian}) corresponds to the planar Gross-Pitaevskii theory, 
\begin{equation}
    H = 2\pi \int_{0}^{\infty} \left[\frac{|\nabla\psi|^{2}}{2m}-\frac{\lambda}{8m^{2}}|\psi|^{4} \right]rdr.
\end{equation}
For arbitrary \(\mu, n\) we obtain that \(H = 0\), which is a consequence of scale invariance and special conformal symmetry. Indeed, the corresponding symmetry generators $D$ and $K$ are
\begin{equation*}
\begin{split}
    & D = 2tH + \frac{i}{2}\int \vec{x}\left(\psi^{\ast}\vec{\nabla}\psi - \psi\vec{\nabla}\psi^{\ast} \right)d^{2}x, \\
    & K = t^{2}H - tD - \frac{m}{2}\int \vec{x}^{2}(\psi^{\ast}\psi) d^{2}x.
\end{split}
\end{equation*}
Their evolution for stationary configurations is described by the following equations \cite{deKok:2007ur}
\begin{equation}
    \frac{dK}{dt}=-t\frac{dD}{dt},\quad \frac{dD}{dt}=2H,
\end{equation}
while the conservation law \(\dot{D}=0\) results in \(H = 0\).

The U(1) Noether charge is \(\mu\)-independent, which is also a result of the scale and conformal invariance.
\begin{equation}
    N = \int_{-\infty}^{\infty} |\psi(t,\vec{x})|^{2} dx^2 =  2\pi \int_{0}^{\infty} \bar{h}(\bar{r})^{2} \bar{r}d\bar{r} = const.
\end{equation}
Thus, the theory describes not a branch of Q-tube solutions with \(H\), \(N\) depending on \(\mu\), but a single point on the plane \(H(N)\) for arbitrary winding number \(n\). 

In the paper \cite{Galushkina:2025yce} we have shown that these results remain relevant if we move on to studying the relativistic generalization of the theory (\ref{NR lagrangian}). One is able to study the influence of conformal symmetry restoration on properties of solitons in the presence of small relativistic corrections. Indeed, let us consider the $(2+1)$-dimensional Lorentz-invariant Lagrangian of a complex scalar field with quartic self-interaction
\begin{equation} 
\label{lagrangian}
    \mathcal{L} = \partial^{\mu}\phi^* \partial_{\mu}\phi- m^2 \phi^* \phi + \frac{\lambda}{2} (\phi^* \phi)^2,
\end{equation}
where the scale invariance and conformal symmetries are broken by the relativistic corrections. To obtain (\ref{NR lagrangian}), one should use the substitution \begin{equation}\label{NR ansatz}
    \phi(t,\vec{x}) = \frac{1}{\sqrt{2m}}e^{-imt}\psi(t,\vec{x})
\end{equation}
and neglect the \(\dot{\psi}^*\dot{\psi}\) term in the resulting Lagrangian.

The Q-tube ansatz is now \begin{equation}
\label{ansatz_background}
    \phi = e^{-{\rm{i}}\omega t} e^{{\rm{i}} n \theta} f(r),
\end{equation}
where \(\omega = m - \mu\). Now, it can be checked that the time dependence of $\psi(t,\vec{x})$ is $\psi(t,\vec{x})\propto e^{-i(\omega-m)t}$. Thus, $|\dot{\psi}|^{2}=|(m-\omega)^{2}\psi^{2}|$ and $|m\psi^{\ast}\dot{\psi}| = |m(m-\omega)\psi^{2}|$. In the limit $\omega \to m$, the ratio \(|\dot\psi|^2 /|m\psi^{\ast}\dot{\psi}|\) is suppressed as $\frac{m-\omega}{m}$. Overall, the non-relativistic limit occurs when $\omega\to m$.

The field equation takes the form \begin{equation}\label{eq}
    f^{''}(r)+\frac{f^{'}(r)}{r} - \frac{n^2}{r^2} f(r) - (m^2 -\omega^2) f(r) + \lambda f^3(r) = 0,
\end{equation}
where \(\omega < m\).

The U(1) Noether charge of the relativistic theory is given by the formula
\begin{equation}
\label{Q}
Q = \int {\rm{i}}\left[ \phi^{*} \partial_0 \phi - \phi \partial_0 \phi^{*}\right] d^2x = 2 \pi \int^{\infty}_{0} 2\omega f^2(r) r dr.
\end{equation}
The energy of soliton \(E\) is provided by the integral
\begin{equation}
\label{E}
\begin{split}
&E = \int \left[\partial_0 \phi^*\partial_0 \phi + \partial_i \phi^*\partial_i \phi + m^2 \phi^*\phi - \frac{\lambda}{2} (\phi^*\phi)^2 \right]d^2x = \\
&\,\,\,\,\,\, 2 \pi \int^{\infty}_{0} \left[(\omega^2 + m^2 + \frac{n^2}{r^2}) f(r)^2 + (f'(r))^2 - \frac{\lambda}{2} f^4(r) \right] r dr.
\end{split}
\end{equation}
Note that the potential in this model does not have a lower bound, which means that it breaks at large values of \(\phi\) because of the sign of the quartic term. In this scope, several UV-completion schemes could be applied to solve this issue. 

 Similarly to (\ref{NR_sc}), we rewrite the field equations in terms of new variables,
\begin{equation}
\label{scaling}
    \begin{split}
        & \tilde{r} = r \sqrt{m^2 - \omega^2} , \\
        & \tilde{f} = f\frac{\sqrt{\lambda}}{\sqrt{m^2 - \omega^2}}.
    \end{split}
\end{equation}
The equations are independent on \(\omega\), \(\lambda\):
\begin{equation}\label{Scaled_eq}
    \tilde{f}^{''}(\tilde{r})+\frac{\tilde{f}^{'}(\tilde{r})}{\tilde{r}} - \left(1+\frac{n^2}{\tilde{r}^2} \right)\tilde{f}(\tilde{r}) + \tilde{f}^{3}(\tilde{r})= 0.
\end{equation}

\begin{figure}[t!]
    \includegraphics[width=0.5\linewidth]{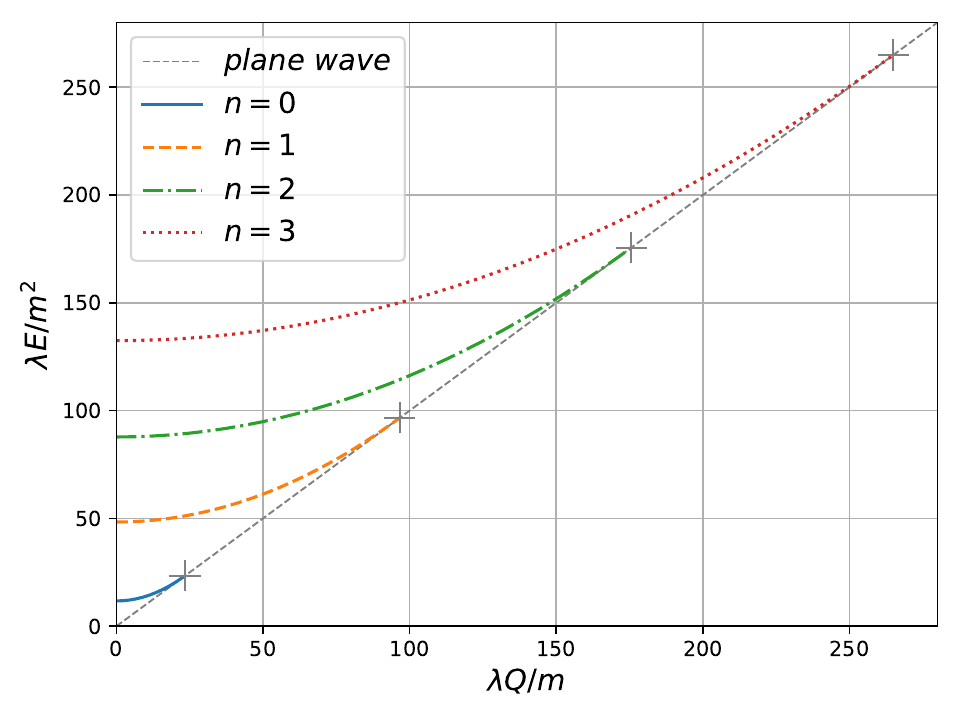}
    \caption{Integral characteristics of Q-tubes with different values of the parameter n.}
    \label{E(Q)_1}
\end{figure}

Using (\ref{scaling}) we can also obtain scaling formulas for \(E\) and \(Q\). First, 
\begin{equation}
\label{Scaled_Q}
    Q = \frac{\omega}{m} Q_{max},
\end{equation}
where
\begin{equation*}\label{Q_max}
    Q_{max} = \frac{4 m \pi}{\lambda} \int^{\infty}_0 \tilde{f}^2(\tilde{r}) \tilde{r} d\tilde{r}
\end{equation*}
is a finite dimensionless constant that can be evaluated numerically. Importantly, \(Q = Q_{max}\) at \(\omega = m\).

Second, in the similar form with explicit dependence on \(\omega\)
\begin{equation}\label{Scaled_E}
    E = \frac{m^2 - \omega^2}{m^2}E_{0} + \frac{\omega^2}{m} Q_{max},
\end{equation}
where
\begin{equation*}\label{E_{max}}
    E_{0} = \frac{2 m^2 \pi}{\lambda} \int^{\infty}_0 \left[ \left(1 + \frac{n^2}{\tilde{r}^2}\right) \tilde{f}^2(\tilde{r}) + (\tilde{f}'(\tilde{r}))^2 - \frac{1}{2} \tilde{f}^4(\tilde{r})\right] \tilde{r} d\tilde{r},
\end{equation*}
which is the energy of the static soliton (\(\omega = 0\)). The condition $d E/d \omega=\omega d Q/d \omega$ results in
\[
2m\omega Q_{max}-2\omega E_0-=m\omega Q_{max}.
\]
Thus, $E_0=m Q_{max}/2$, and the \(E(Q)\) dependence is determined by single quantity that should be obtained numerically. The results for integral characteristics are given in Fig.\ref{E(Q)_1}.

As one can see, the maximal values of \(E\) and \(Q\) are obtained at \(\omega = m\). In this point, \(Q = Q_{max} = N\), where \(N\) is a Noether charge for the corresponding winding number in the non-relativistic conformal field theory. Similarly, \(E_{max} - m Q_{max} = H = 0\), where H is the Hamiltonian of the non-relativistic CFT. 

According to the Vakhitov-Kolokolov instability criterion 
\begin{equation}
    \frac{\omega}{Q}\frac{dQ}{d\omega}>0,
\end{equation}
the solutions for arbitrary \(n\) are unstable. This result can also be obtained by considering the criterion of kinematic stability. For these solitons \(E > mQ\), which means that they fission quantum-mechanically into free particles.

However, there are several ways of UV-completion of the theory (\ref{lagrangian}) which allows one to obtain a branch of stable solutions and does not break the Schr\"{o}dinger group in the NR-limit of the theory.

\section{Solitons in the model with an additional sextic term}\label{section 2}

The energy functional in model (\ref{lagrangian}) has no lower bound, thus it is rational to seek for options to UV-complete the theory. A straightforward approach is to add a higher-order $U(1)$-invariant term to the potential. Thus, the potential becomes bounded from below.
In particular, our first choice of the UV-completion is described by the Lagrangian (originally considered in \cite{Coleman:1985ki})
\begin{equation} 
\label{lagrangian_6}
    \mathcal{L} = \partial^{\mu}\phi^* \partial_{\mu}\phi- m^2 \phi^* \phi + \frac{\lambda}{2} (\phi^* \phi)^2 - \frac{\sigma}{3} (\phi^* \phi)^3,
\end{equation}
where \([m] = M\), \([\lambda] = M\), \([\sigma] = 1\); \(\lambda, \sigma > 0\). 
At the regime \(|\phi|^2\ll \lambda/\sigma\) this model approximately coincides with (\ref{lagrangian}).

The theory (\ref{lagrangian_6}) supports condensate solutions. Using the ansatz $\psi(t,\vec{x}) = F e^{i\omega t} $, where \(F\) is a constant, we obtain a spatially homogeneous solution. According to the equations of motion, for bound states with \(\omega < m\),
\begin{equation}\label{condensate_6}
      F = \pm \sqrt{\frac{\lambda - \sqrt{\lambda^2 - 4(m^2 - \omega^2) }}{2 \sigma}}.
\end{equation} 
The existence of this condensate will be important to interpret the behavior of solitons in this modified model.

Now we are going to study non-topological solitons. Following the ansatz given in Eq.(\ref{ansatz_background}), we obtain the field equation for solitons
\begin{equation}\label{eq_6}
    f^{''}(r)+\frac{f^{'}(r)}{r} - \frac{n^2}{r^2} f(r)- (m^2 -\omega^2) f(r) + \lambda f^3(r) - \sigma f^5(r) = 0,
\end{equation}
where \(\omega < m\).

The expression for U(1) Noether charge is similar as in the previous model
\begin{equation}
\label{Q_6}
Q = 2 \pi \int^{\infty}_{0} 2\omega f^2(r) r dr.
\end{equation}
The energy of soliton \(E\) is now
\begin{equation}
\label{E_6}
\begin{split}
&E = 2 \pi \int^{\infty}_{0} \left[(\omega^2 + m^2 + \frac{n^2}{r^2}) f(r)^2 + (f'(r))^2 - \frac{\lambda}{2} f^4(r) + \frac{\sigma}{3} f^6(r) \right] r dr.
\end{split}
\end{equation}

Now we rescale the functions, coordinates and parameters. In this model, it is impossible to get rid of \(\omega\), as it was done in (\ref{eq}). To reduce the number of parameters, we use the scaling which differs from (\ref{scaling}):
\begin{equation}
\label{scaling_6}
         \mathrm{w} = \frac{\omega}{m}, \quad
         \tilde{r} = mr, \quad
         \tilde{f} = f\frac{\sqrt{\lambda}}{m}, \quad
         g = \sigma \frac{m^2}{\lambda^2}.
\end{equation}

The field equation takes the form
\begin{equation}\label{Scaled_eq_6}
    \tilde{f}^{''}(\tilde{r})+\frac{\tilde{f}^{'}(\tilde{r})}{\tilde{r}} - \frac{n^2}{\tilde{r}^2} \tilde{f}(\tilde{r}) - \left(1 - \mathrm{w}^2 \right)\tilde{f}(\tilde{r}) + \tilde{f}^{3}(\tilde{r}) - g \tilde{f}^{5}(\tilde{r}) = 0.
\end{equation}

In order to examine the regimes of solutions at different \(g\) and \(\mathrm{w}\), one can consider the mechanical interpretation of the equation (\ref{Scaled_eq_6}), where \(\tilde{r}\) is a time coordinate, \(\tilde{f}\) is a dynamic coordinate, \(\tilde{f}^{'}(\tilde{r})/\tilde{r}\) is a 'time'-dependent friction term and 
\begin{equation}\label{U_6}
    U(\tilde{f}) = - \frac{n^2}{2\tilde{r}^2} \tilde{f}^2 - \frac{(1 - \mathrm{w}^2)}{2} \tilde{f}^2 + \frac{\tilde{f}^4}{4} - g \frac{ \tilde{f}^6}{6} 
\end{equation}
is a 'time'-dependent mechanical potential. However, in the case of \(n=0\) the potential is static. 

\begin{figure}[H]
\centering
\subfloat{\includegraphics[width=0.5\linewidth]{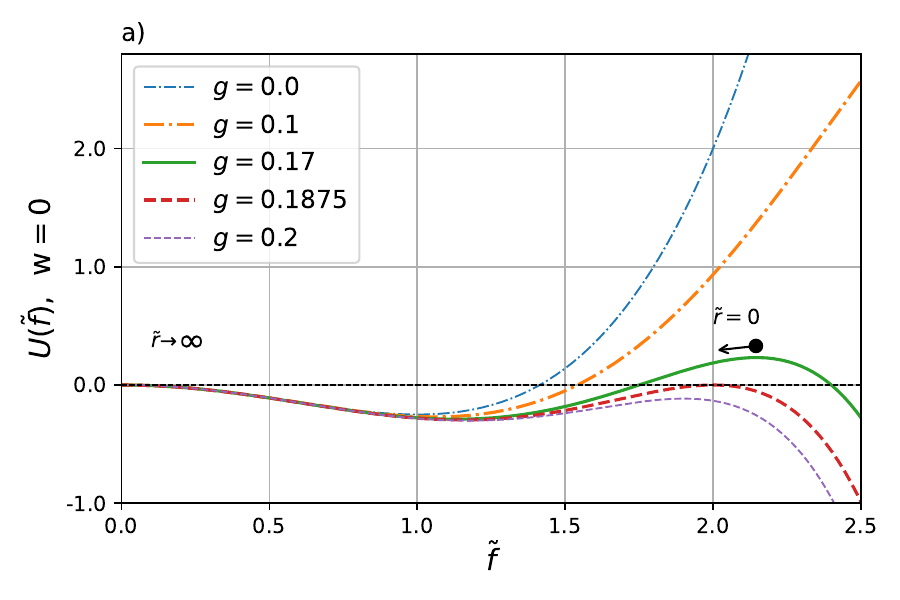}}
\subfloat{\includegraphics[width=0.5\linewidth]{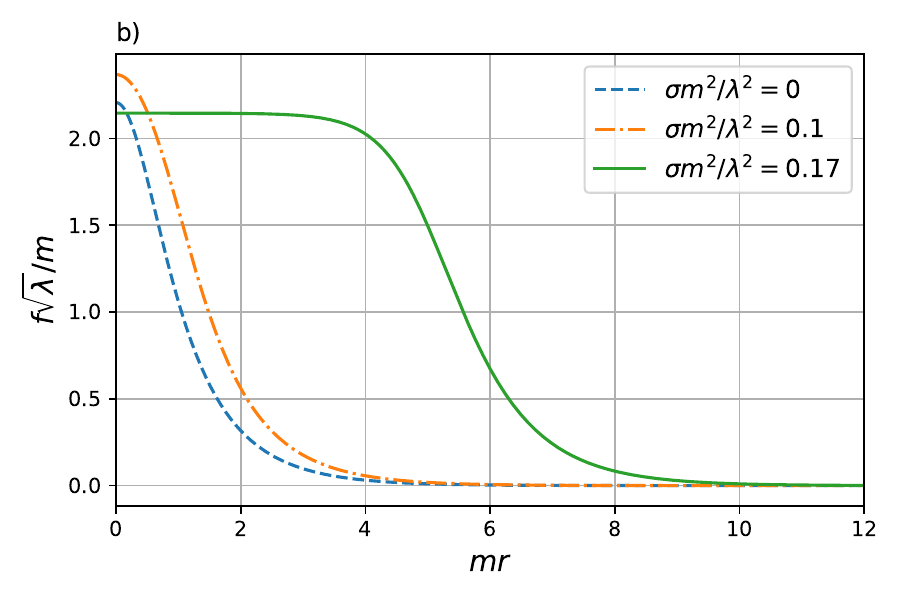}}
\caption{a) The mechanical potential \(U(\tilde{f})\) for different values of \(g\); \(n = 0\), \(\mathrm{w} = 0\). b) Profiles of solitons for different values of \(\sigma m^2 / \lambda^2\); \(n = 0\), \(\mathrm{w} = 0\).}
    \label{V_6}
\end{figure}

In this terms, the condensate (\ref{condensate_6}) corresponds to the position of a particle, laying on the maximum of the potential curve (see Fig. \ref{V_6}). The soliton solution is interpreted as a trajectory of the moving particle, starting at some value of the dynamic coordinate \(\tilde{f}\) at the initial moment (\(\tilde{r} = 0\)) and ending at the origin at \(\tilde{r} = \infty\) (e. g. see Fig. \ref{V_6}.a). If the particle trajectory starts from the maximum of the potential curve (e. g. for \(g = 0.17\)), the corresponding soliton is in the thin-wall regime (see Fig. \ref{V_6}.b), which means that it has a core consisting of the condensate (\ref{condensate_6}). The trajectory starting from the slope of the potential curve (e. g. for \(g = 0\)) corresponds to the soliton without a condensate core. 

\begin{figure}[H]
\centering
\subfloat{\includegraphics[width=0.5\linewidth]{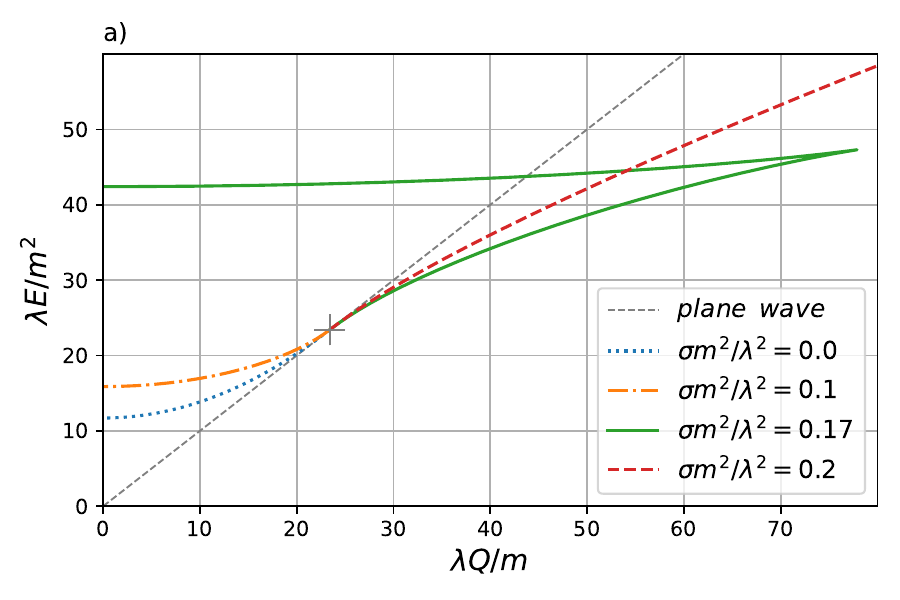}}
\subfloat{\includegraphics[width=0.5\linewidth]{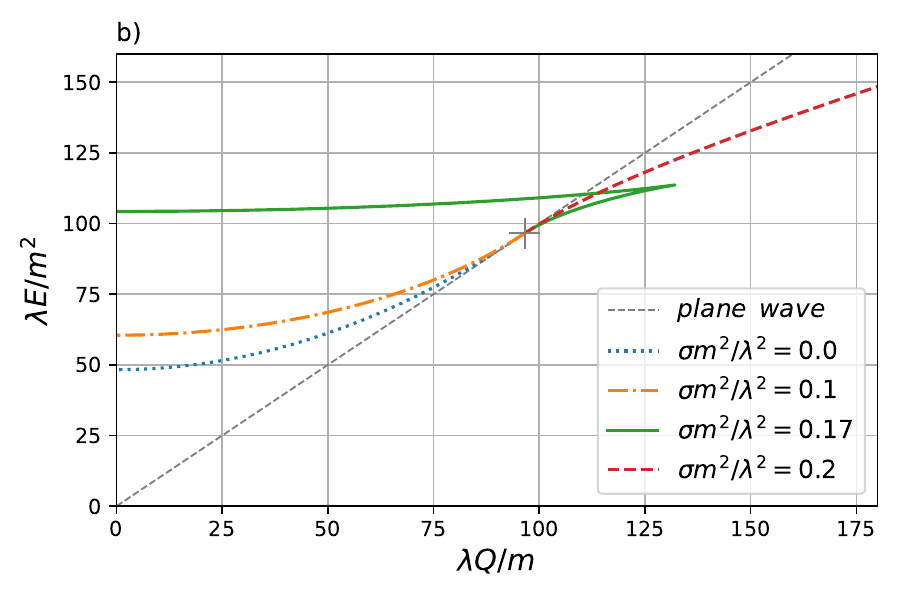}}\\
\caption{Integral characteristics of Q-tubes with different values of the parameter \(\sigma m^2 / \lambda^2 \): a) \(n = 0\), b) \(n = 1\).}
    \label{E(Q)_6}
\end{figure}

The frequency of the soliton (\(\omega = m \mathrm{w} \)) is bound by the inequality \(\mathrm{w}_{min} < \mathrm{w} < 1\). For solitons without a condensate core, the lower bound is trivial, \(\mathrm{w}_{min} = 0\), while for thin-wall solitons the value of \(\mathrm{w}_{min}\) can be found in terms of mechanical interpretation. Indeed, as one can see in Fig. \ref{V_6}, for \(\mathrm{w} = 0\) the global maximum of potential curve for \(g > 3 / 16\) is reached at \(\tilde{f} = 0\). In other words, while $g\leq 3/16$ the state $|\phi|=0$ is a meta-stable vacuum and there is a kinematically unstable soliton with $Q=0$ and $E>0$. For $g>3/16$, the particle sliding from the condensate point cannot reach the position \(\tilde{f} = 0\), so there are no solitons at \(\mathrm{w} = 0\), and \(\mathrm{w}_{min}\) is nontrivial. The value \(\mathrm{w}_{min}\) corresponds to the potential curve with two global maxima, and at both of them \(U = 0\). Thus, we obtain that  
\begin{equation*}
    \mathrm{w}^2_{min} = 1 - \frac{3}{16 g}.
\end{equation*}

The properties of solitons depend on the value of \(\sigma = g \lambda^2 / m^2\). In particular, their stability is determined by this parameter. To demonstrate this, we solve the equation (\ref{Scaled_eq_6}) numerically and consider the integral characteristics (\ref{Q_6}, \ref{E_6}) of the solutions. The results are shown in Fig.\ref{E(Q)_6}.a. For this model, there are tree type of \(E(Q)\) curves. One of them (\(\sigma m^2 / \lambda^2 = 0.1\)) corresponds to an unstable soliton branch without a thin-wall regime, similar to one in the purely quartic model (\(\sigma m^2 / \lambda^2 = 0\)). For \(\sigma m^2 / \lambda^2 = 0.17\) we obtain a branch of solitons with kinematically stable relativistic solutions (\(E < mQ\)), reaching the thin-wall limit when \(\omega\) goes to zero. Finally, for \(\sigma m^2 / \lambda^2 = 0.2\) all the solutions are kinematically stable, and there are thin-wall solitons at \(\omega \to \omega_{min}\). Their charge and energy have no upper bound.

\begin{figure}[h!]
    \includegraphics[width=0.5\linewidth]{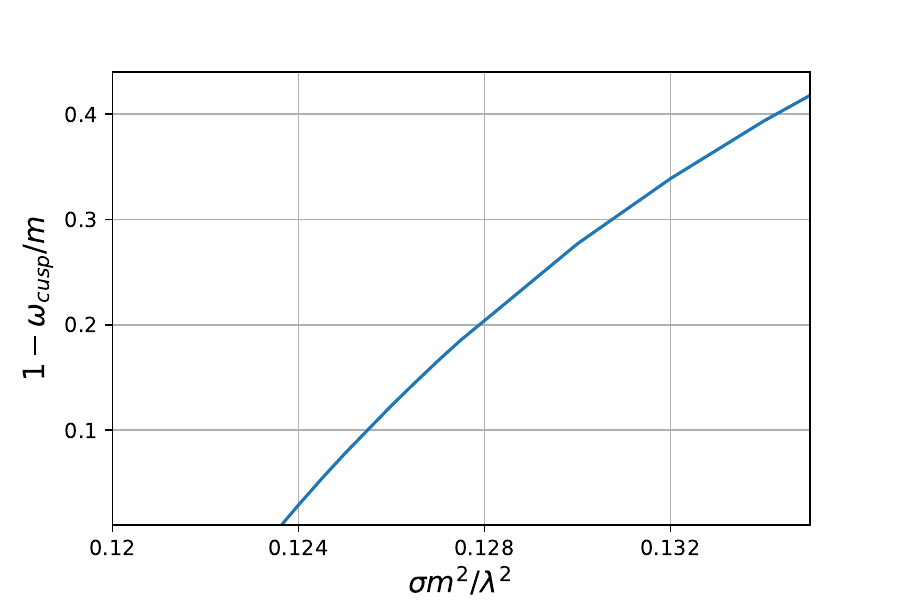}
    \caption{The value \(1 - \omega_{cusp} / m\) for Q-tubes with different values of the parameter \(\sigma m^2 / \lambda^2 \); n = 0.}
    \label{omega_6_cusp}
\end{figure}

The cusp of \(E(Q)\) curve exists for \(0.124 \lesssim \sigma m^2 / \lambda^2 < 3/16\). The cusp is reached at some value of the frequency parameter, \(\omega = \omega_{cusp}\). The upper limit is determined by the existence of nontrivial \(\omega_{min}\), while the lower bound is found numerically (see Fig. \ref{omega_6_cusp}).

The similar types of branches characterize solutions with \(n = 1\) (see Fig.\ref{E(Q)_6}.b).

One can note that, both for \(n = 0\) and \(n = 1\), all the branches end in the same point which was previously described as a conformal limit of a theory with quartic self-interaction. This is because \(|\phi|^2\ll \lambda/\sigma\). The field amplitude vanishes in the limit \(\omega \to m\). However, it is quite surprising that soliton branches coinciding at \(\omega = m\) have such different stability properties in the vicinity of this point.

\section{Q-tubes in the Friedberg-Lee-Sirlin  model}\label{section 3}

\begin{figure}[h!]
    \includegraphics[width=1.\linewidth]{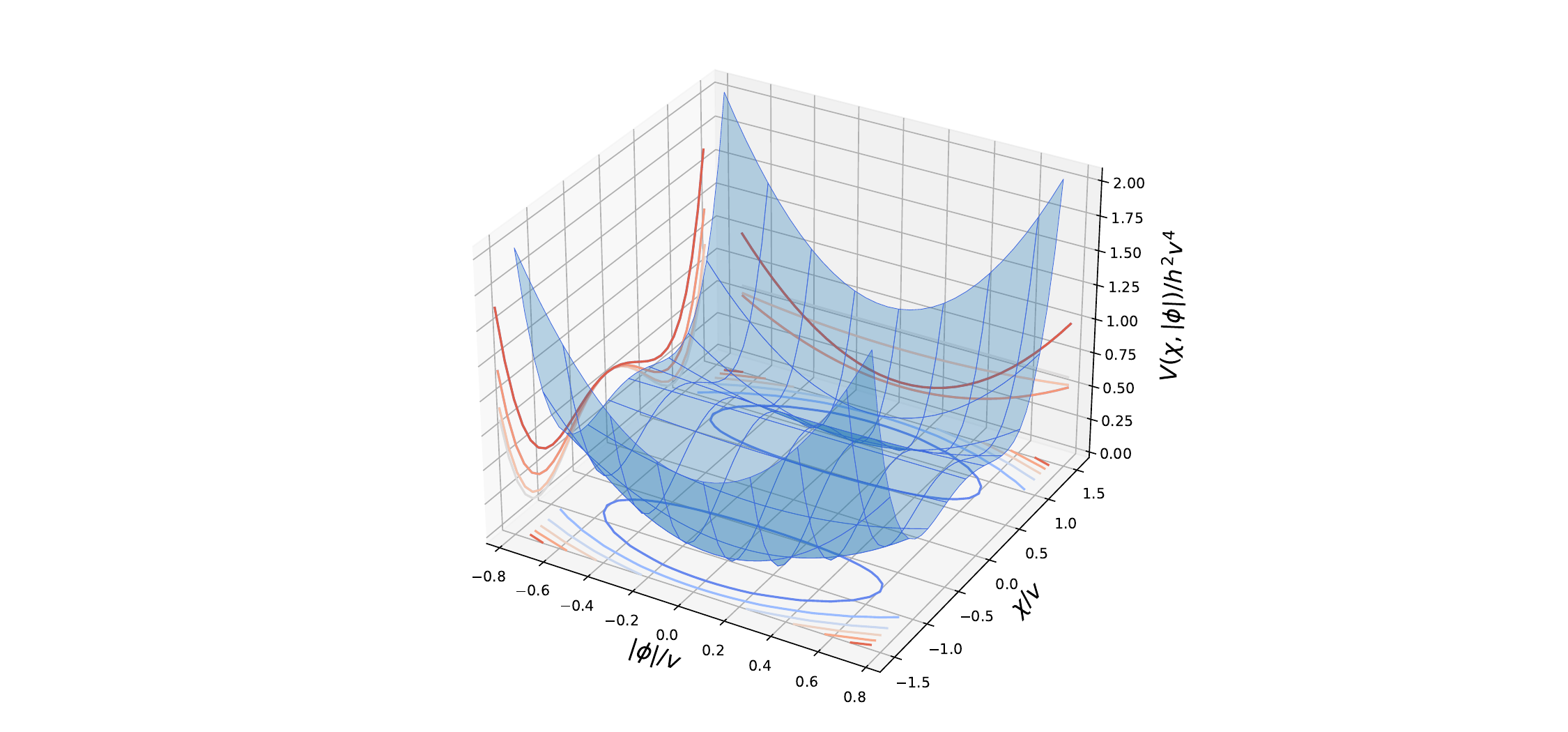}
    \caption{The potential of FLS theory for \(\varkappa/h = 1\).}
    \label{2D_1}
\end{figure}

Now let us consider Friedberg-Lee-Sirlin (FLS) model \cite{Friedberg:1976me}, which can reproduce the theory (\ref{lagrangian}) at low energies \cite{Kim:2023zvf}. The Lagrangian is 
\begin{equation}\label{FLS_Lagrangian}
            \mathcal{L} = \partial_{\mu}\phi^{\ast}\partial^{\mu}\phi + \frac{1}{2}\partial_{\mu}\chi\partial^{\mu}\chi - V(\chi, |\phi|),
    \end{equation}
where 
\begin{equation}\label{FLS_potential}
            V(\chi, |\phi|) = h^{2}|\phi|^{2}\chi^{2} + \frac{ \varkappa^{2} }{2}\left(\chi^{2} - v^{2} \right)^{2},
    \end{equation}
and \([v] = [h] = [\varkappa] = M^{1/2}\). This theory allows for spontaneous symmetry violation of (\(\chi \to - \chi\)). The complex field \(\phi\) acquires mass due to the nontrivial vacuum of the real field (\(\chi = v\)), \(m_{\phi} = hv\). The potential of FLS theory for \(\varkappa/h = 1\) is shown in Fig.\ref{2D_1}. As one can see, it possesses a flat direction at $\chi=0$ axis (the field $\phi$ becomes massless).

\begin{figure}[h!]
    \includegraphics[width=0.5\linewidth]{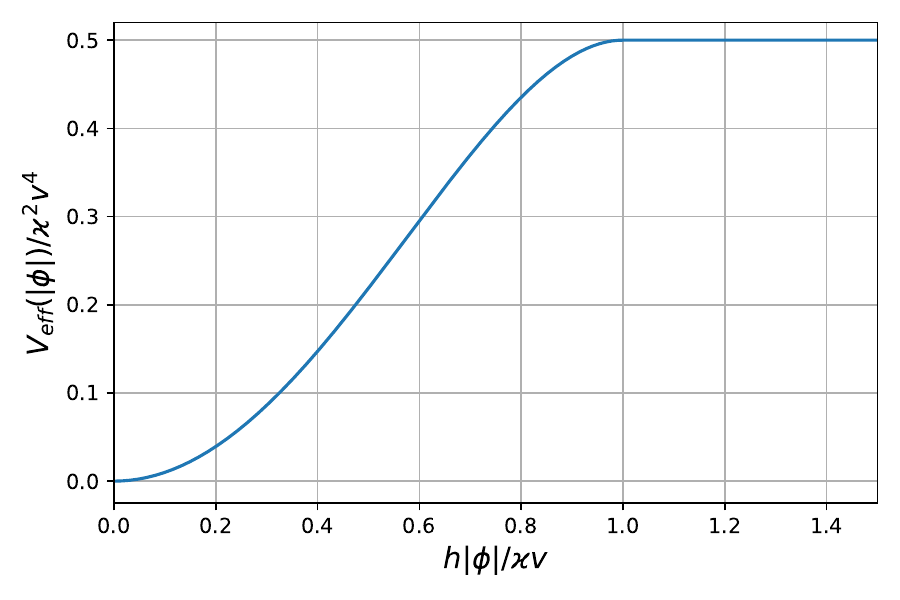}
    \caption{The effective potential \(V_{eff}(|\phi|)\).}
    \label{EFF}
\end{figure}

The equations of motion for the theory (\ref{FLS_Lagrangian}) are
\begin{equation}\label{FLS_eq}
    \begin{split}
        &\partial_{\mu}\partial^{\mu}\phi + h^{2}\chi^2\phi = 0,\\
        &\partial_{\mu}\partial^{\mu}\chi + 2h^{2}|\phi|^{2}\chi + 2\varkappa^{2}(\chi^{2}-v^{2})\chi = 0.
    \end{split}
\end{equation}
In the limit \(m_{\chi} = \varkappa v \gg m_{\phi}\) the kinetic term for the field \(\chi\) can be neglected, and we obtain that \(\chi^2 = v^2 - \frac{h^2}{\varkappa^2}|\phi|^{2}\) or \(\chi = 0\). In this case, we can consider an effective theory of a massive self-interacting field \(\phi\) with a piece-wise potential \cite{Kim:2023zvf}

\begin{equation}\label{Eff_potential}
  V_{eff}(|\phi|^{2})=\begin{cases}
    m_{\phi}^{2}|\phi|^2-\frac{h^4}{2\varkappa^2}|\phi|^4, & \text{if $|\phi|<\frac{\varkappa v}{h}$},\\
    \frac{\varkappa^2v^4}{2}, & \text{if $|\phi|>\frac{\varkappa v}{h}$}.
  \end{cases}
\end{equation}
As one can see, for small \(\phi\) this theory coincides with the theory (\ref{lagrangian}). The potential (\ref{Eff_potential}) is plotted in Fig. \ref{EFF}.

Now let us consider Q-tube solutions in FLS theory with the following ansatz
\begin{equation}\label{FLS_ansatz}
\begin{cases}
    &\phi(t,\vec{x}) = e^{-{\rm{i}}\omega t} e^{{\rm{i}} n \theta} f(r),\\ &\chi(t,\vec{x})=g(r).
\end{cases}
\end{equation}

Equations on these functions are
\begin{equation}\label{FLS_Q_tube}
    \begin{split}
        &f^{''}+\frac{f^{'}}{r} - \frac{n^2}{r^2} f - (h^2 g^2-\omega^2) f = 0,\\
        &g^{''}+\frac{g^{'}}{r} - 2h^{2}f^{2}g - 2\varkappa^{2}(g^{2}-v^{2})g = 0.
    \end{split}
\end{equation}

The expression for U(1) Noether charge is similar to (\ref{Q}), and the energy is written as
\begin{equation}
\label{E_FLS}
\begin{split}
&E = \int \left[\partial_0 \phi^*\partial_0 \phi + \partial_i \phi^*\partial_i \phi + \frac{1}{2}\partial_0 \chi^*\partial_0 \chi + \frac{1}{2} \partial_i \chi^*\partial_i \chi + h^{2}|\phi|^{2}\chi^{2} + \frac{ \varkappa^{2} }{2}\left(\chi^{2} - v^{2} \right)^{2} \right]d^2x = \\
&\,\,\,\,\,\, 2 \pi \int^{\infty}_{0} \left[(\omega^2 + \frac{n^2}{r^2}) f(r)^2 + (f'(r))^2 + \frac{1}{2} (g'(r))^2 +  h^{2}f(r)^{2}g(r)^{2} + \frac{ \varkappa^{2} }{2}\left(g(r)^{2} - v^{2} \right)^{2} \right] r dr.
\end{split}
\end{equation}
We can rescale variables and parameters:
\begin{equation}
\label{scaling_FLS}
         \mathrm{w} = \frac{\omega}{hv}, \quad
         \tilde{r} = hvr, \quad
         \tilde{f} = \frac{f}{v}, \quad
         \tilde{g} = \frac{g}{v}, \quad
         \tilde{\varkappa} = \frac{\varkappa}{h}.
\end{equation}
The equations of motion take the form
\begin{equation}
    \begin{split}
        &\tilde{f}^{''}+\frac{\tilde{f}^{'}}{\tilde{r}} - \frac{n^2}{\tilde{r}^2} \tilde{f} - (\tilde{g}^2-\mathrm{w}^2) \tilde{f} = 0,\\
        &\tilde{g}^{''}+\frac{\tilde{g}^{'}}{\tilde{r}} - 2 \tilde{f}^{2}\tilde{g} - 2\tilde{\varkappa}(\tilde{g}^{2}-1)\tilde{g}= 0.
    \end{split}
\end{equation}

\begin{figure}[H]
\centering
\subfloat{\includegraphics[width=0.5\linewidth]{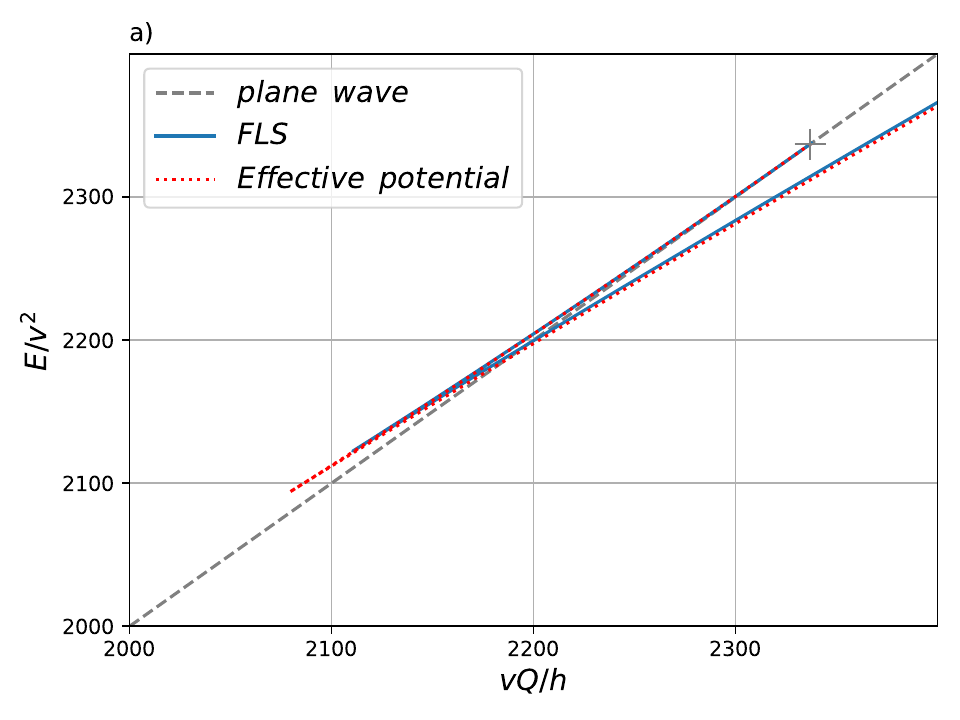}}
\subfloat{\includegraphics[width=0.5\linewidth]{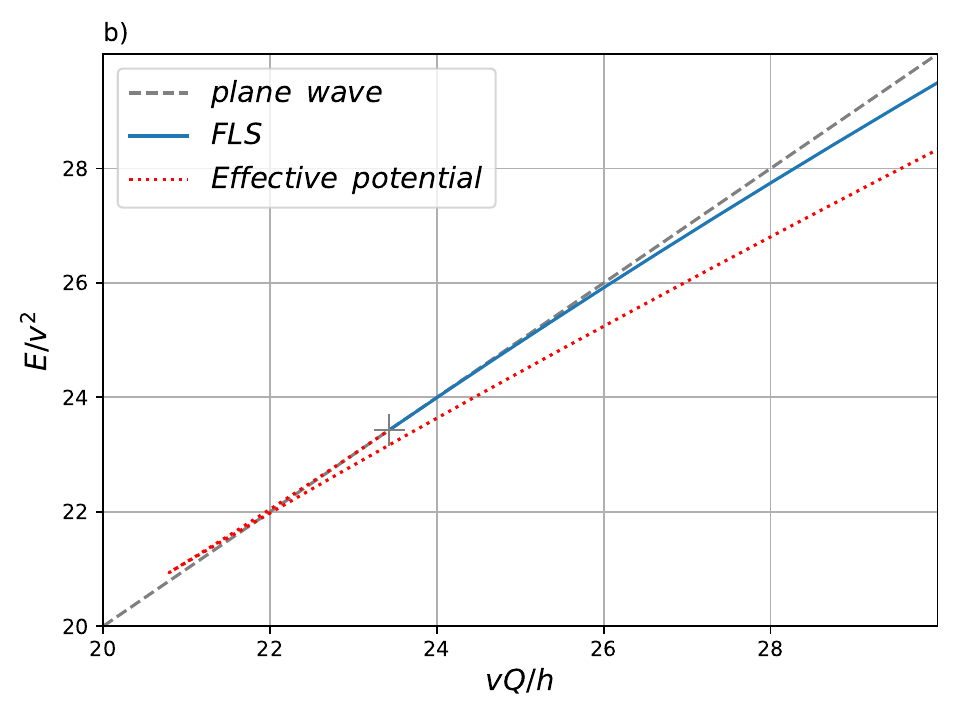}}
\caption{Integral characteristics of Q-tubes with n = 0: a) \(\varkappa / h = 0.1\), b) \(\varkappa / h = 1\).}
    \label{FLS_10}
\end{figure}

We solve these equations numerically and obtain integral characteristics of solitons. In the case of FLS theory, there are two types of \(E(Q)\) curves. The first of them (see Fig. \ref{FLS_10}.a), corresponds to the small values of \(\tilde{\varkappa}\). This curve starts as a branch of kinematically unstable solitons (\(E > hvQ\)), but then, after a cusp, goes to the kinematically stable area. For \(\tilde{\varkappa} \ll 1\) the approximation (\ref{Eff_potential}) is valid, and we can compare the result of calculations for FLS theory with the result for effective theory of complex scalar field. Both curves end in the point where the conformal symmetry is restored, as soon as they can be treated independently as a UV-completion of the theory (\ref{lagrangian}).

\begin{figure}[h!]
    \includegraphics[width=0.5\linewidth]{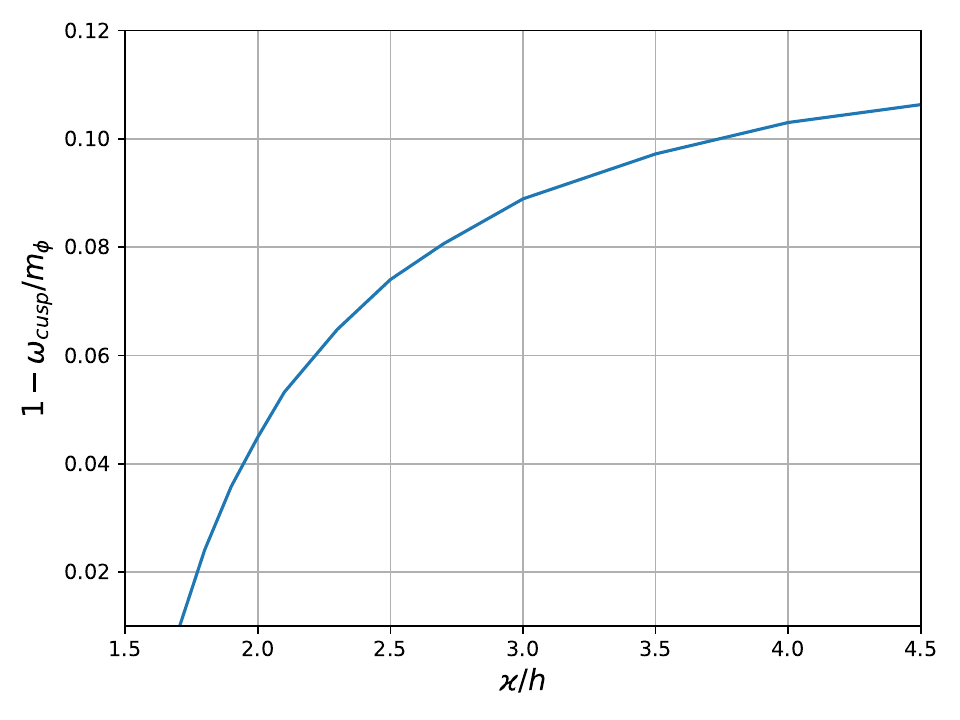}
    \caption{The value \(1 - \omega_{cusp} / m_{\phi}\) for Q-tubes with different values of the parameter \(\varkappa / h\); n = 0.}
    \label{omega_cusp_FLS}
\end{figure}

The cusp vanishes at \(\tilde{\varkappa} \approx 1.7\), see Fig. \ref{omega_cusp_FLS}. In this point, \(1 - \omega_{cusp}/m_{\phi} = 0\).

For larger values of \(\tilde{\varkappa}\), there is no cusp on the \(E(Q)\) curve, and all the solitons are kinematically stable (see Fig. \ref{FLS_10}.b). The approximation (\ref{Eff_potential}) does not represent the properties of \(E(Q)\) curve. However, both curves also end in the point of the conformal symmetry restoration (for more details see Appendix \ref{Appendix}). 

\begin{figure}[H]
\centering
\subfloat{\includegraphics[width=0.5\linewidth]{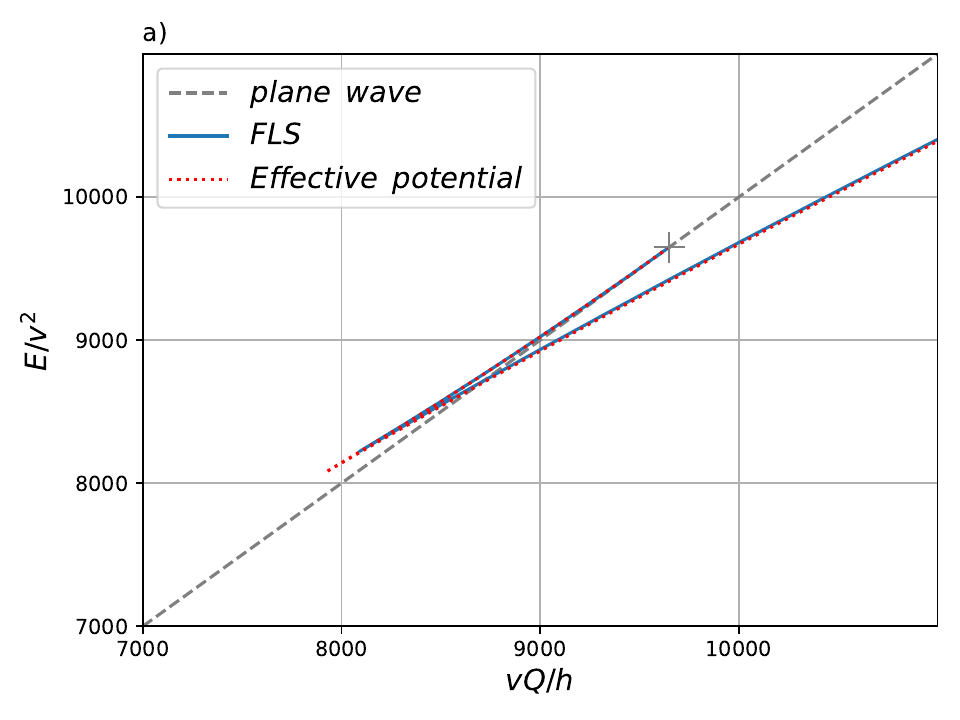}}
\subfloat{\includegraphics[width=0.5\linewidth]{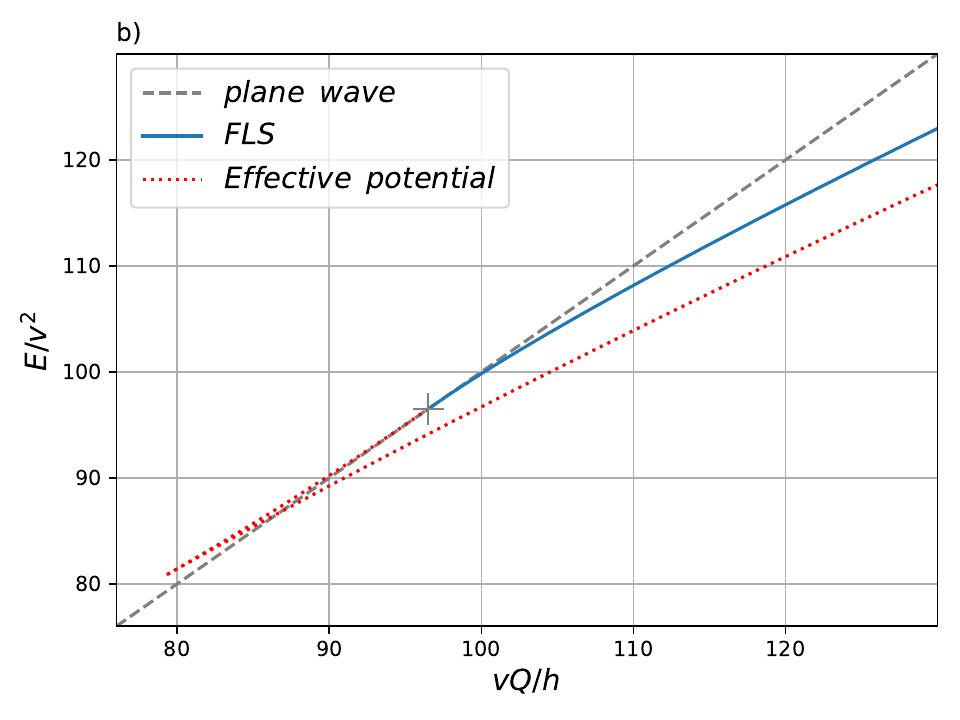}}
\caption{Integral characteristics of Q-tubes with n = 1: a) \(\varkappa / h = 0.1\), b) \(\varkappa / h = 1\).}
    \label{FLS_10_1}
\end{figure}

Our results indicate that for Q-tubes with higher winding numbers $n\geq 1$, the results are qualitatively similar to those found for $n=0$, see Fig. \ref{FLS_10_1}.

Thus, we can conclude that, similarly to the theory (\ref{lagrangian_6}), for the FLS theory there are stable branches of solitons with conformal symmetry restoration in the non-relativistic limit.

\section{Outlook}
In this paper, we considered several Lorentz-invariant scalar field theories in $(2+1)$ dimensions which support non-topological solitons. At the low-energy limit of these theories, there is a conformal symmetry restoration, which determines the properties of non-topological solitons. 
First, to illustrate this mechanism, we considered the non-relativistic theory with an unbroken conformal symmetry and then discussed its relativistic generalization, as it was done in \cite{Galushkina:2025yce}. The solitons in this model are unstable. Then, in the similar way, we studied solitons in the theory with an additional sextic self-interaction. This theory can be considered as a mechanism of UV-completion for the previous theory. The stability of the solutions is changed by the additional nonlinearity, and we obtain stable branches of solitons with the thin-wall regime. At low energies, the conformal symmetry is restored in the same way as in the theory with pure quartic self-interaction. 

Finally, we considered non-topological solitons in $(2+1)$-dimensional Friedberg-Lee-Sirlin theory, which can also be considered as a UV-completion of the theory with an attractive quartic self-interaction. The study of the non-relativistic limit for the theory of two scalar fields is more complicated than in the case of one field, but the results are similar to ones for the theory with additional sextic term (however, there are no thin-wall solitons in FLS model). We have shown that the conformal symmetry is restored in the non-relativistic limit of FLS theory. We obtained stable branches of solitons with conformal symmetry restoration. The presence of an unbroken Schr\"{o}dinger group may significantly alter the mechanism of soliton production in the early Universe \cite{Krylov:2013qe, kasuya2025chargedistributiongaugemediationtype}.

In nonlinear theories that we present in this paper, a different class of solutions arises -- self-similar configurations \cite{sulem2007nonlinear}. Indeed, the study of the nonlinear Schr\"{o}dinger equation indicates that symmetry transformations for wave function might generate blow-up solutions. Remarkably, this blow-up instability leads to a rapid increase of density at finite times. These solutions may be of particular interest in the DM phenomenology \cite{dmitriev2025selfsimilarkineticsgravitationalboseeinstein}. Moreover, in a model of ultra-light dark matter considered as axion-like particles, the planar dynamics is described by nonlinear Schr\"{o}dinger equation \cite{Brax:2025uaw}. Thus, one may use blow-up solutions to estimate the typical time of production of dense configurations, which is crucial for direct axion searches \cite{Tinyakov:2015cgg}.  


Following original paper \cite{zel1971generation}, one can expect similar superradiance effect for our spinning solutions. In the case of Q-balls, this effect was studied in \cite{Saffin:2022tub,Zhang:2024ufh}. It was established that superradiance can occur even for spatially non-rotating field configurations due to the spinning in the internal $U(1)$ space. The recent research \cite{zhang2025qballsuperradianceanalyticalapproach} provides analytical approach that simplifies the study of superradiance amplification factors.

\section*{Acknowledgments}

Numerical studies of spinning non-topological solitons were supported by the grant RSF 22-12-00215-$\Pi$. The work of E. Kim was supported by the Foundation for the Advancement of Theoretical Physics and Mathematics BASIS.

\appendix

\section{NR limit and gradient approximation}
\label{Appendix}

In this appendix, we show that it is of no surprise for the conformal symmetry restoration to happen in the same isolated point for FLS model and our EFT. As a first step, we assume gradient approximation for "heavy" field $\chi$ in the limit $\omega\to hv$:
\begin{equation}\label{grad.approx}
   \frac{\partial_{\mu}\partial^{\mu}\chi}{\kappa^{2}v^{2}\chi}\to0 : \chi^{2}=v^{2}-\frac{h^{2}}{\kappa^{2}}|\phi|^{2}. 
\end{equation}
In this limit, both fields of the FLS model tend to their vacuum values as can be seen by studying condensate solutions in FLS model with ansatz (\ref{FLS_ansatz}). Thus,
\begin{equation*}
    \begin{cases}
        &\phi = 0 + \delta\phi; \\
        &\chi = v + \delta\chi.        
    \end{cases}
\end{equation*}
Now, one can show using Eq.(\ref{grad.approx}) that if field $\chi$ tends to $v$ and $\delta \chi\propto \epsilon^{2}$, where $\epsilon\to0$, then $\delta \phi \propto \epsilon$. As a next step, it is crucial to verify that in the limit $\omega \to hv$ gradient approximation is valid. In accordance with written above, let us consider linearized (in powers of $\epsilon$) equations of motion
\begin{equation}\label{lin.n.vac.}
    \begin{split}
        & \nabla_{r}^{2}\delta\phi = (h^{2}v^{2}-\omega^{2})\delta\phi, \\
        & \nabla_{r}^{2}\delta\chi = 2h^{2}v|\delta\phi|^{2} + 4\kappa^{2}v^{2}\delta\chi=0.
    \end{split}
\end{equation}
From the first equation in Eqs.(\ref{lin.n.vac.}) we extract that $\delta\phi\propto e^{-r/L_{\phi}}$, where $L_{\phi}=1/\sqrt{h^{2}v^{2}-\omega^{2}}$. In the gradient approximation, the second equation becomes algebraic with following solution
\begin{equation*}
    \delta\chi = -\frac{h^{2}}{2\kappa^{2}v}|\delta\phi|^{2}.
\end{equation*}
Thus, we conclude that 
\begin{equation}
    \frac{\nabla_{r}^{2}\delta\chi}{m_{\chi}^{2}v} \propto \frac{1}{(L_{\phi}m_{\chi})^{2}}\frac{\delta \chi}{v}, 
\end{equation}
which indicates that in the non-relativistic limit ($\omega\to hv$) the gradient approximation is valid and the isolated point of the FLS model is the same as in theory with effective potential (\ref{Eff_potential}).

\bibliography{biblio}

\end{document}